\documentclass[12pt]{iopart}

\usepackage{graphicx}
\usepackage{hyperref}

\usepackage{amssymb}
\def\Tr{\mbox{Tr}\,}

\newcommand{\I} {\mbox{Im}\,}

\def\Xint#1{\mathchoice
   {\XXint\displaystyle\textstyle{#1}}%
   {\XXint\textstyle\scriptstyle{#1}}%
   {\XXint\scriptstyle\scriptscriptstyle{#1}}%
   {\XXint\scriptscriptstyle\scriptscriptstyle{#1}}%
   \!\int}
\def\XXint#1#2#3{{\setbox0=\hbox{$#1{#2#3}{\int}$}
     \vcenter{\hbox{$#2#3$}}\kern-.5\wd0}}

\def\dashint{\Xint-}

\newcommand{\la}{\label}
\newcommand{\be}{\begin{equation}}
\newcommand{\ee}{\end{equation}}
\newcommand{\bea}{\begin{eqnarray}}
\newcommand{\eea}{\end{eqnarray}}

\newcommand{\p}{\partial}
\newcommand{\ba}{\begin{align}}
\newcommand{\ea}{\end{align}}
\newcommand{\1}{\frac{1}{2}}

\def\Xint#1{\mathchoice
   {\XXint\displaystyle\textstyle{#1}}%
   {\XXint\textstyle\scriptstyle{#1}}%
   {\XXint\scriptstyle\scriptscriptstyle{#1}}%
   {\XXint\scriptscriptstyle\scriptscriptstyle{#1}}%
   \!\int}
\def\XXint#1#2#3{{\setbox0=\hbox{$#1{#2#3}{\int}$}
     \vcenter{\hbox{$#2#3$}}\kern-.5\wd0}}

\def\dashint{\Xint-}

\begin{document}

\today \hfill 

\title[Soliton solutions in hCM \ldots]{Soliton solutions of Calogero model in harmonic potential.
 }

\author{Alexander G. Abanov
}
\address{Department of Physics and Astronomy,
Stony Brook University,  Stony Brook, NY 11794-3800.}

\author{Andrey Gromov
}
\address{Department of Physics and Astronomy,
Stony Brook University,  Stony Brook, NY 11794-3800.}

\author{Manas Kulkarni
}
\address{Department of Physics and Astronomy,
Stony Brook University,  Stony Brook, NY 11794-3800.}

\address{}

\address{Department of Condensed Matter Physics and Material Science,
Brookhaven National Laboratory, Upton, NY-11973. }


\begin{abstract}
A classical Calogero model in an external harmonic potential is known to be integrable for any number of particles. We consider here reductions which play a role of ``soliton'' solutions of the model. We obtain these solutions both for the model with finite number of particles and in a hydrodynamic limit. In the latter limit the model is described by hydrodynamic equations on continuous density and velocity fields. Soliton solutions in this case are finite dimensional reductions of the hydrodynamic model and describe the propagation of lumps of density and velocity in the nontrivial background.
\end{abstract}

\maketitle


\tableofcontents





\section{Introduction}

The harmonic Calogero model (hCM)
\cite{Calogero-1969,Sutherland-1971} describes one-dimensional particles moving in the presence of an external harmonic potential and interacting through
an inverse-square potential.
The Hamiltonian of the model reads
\bea
    {\cal H}_{hCM} &=& \frac{1}{2}\sum_{j=1}^{N}\left(p_j^{2}+\omega^{2}x_{j}^{2}\right)
    +\1 \sum_{j,k=1; j\neq k}^{N} \frac{g^{2}}{(x_{j}-x_{k})^{2}},
  \la{hCM} \\
	&=&  \1 \sum_{j=1}^N \left| p_j- i\omega x_j
	+ ig \sum_{k=1; k\neq j}^{N}\frac{1}{x_j-x_k}  \right|^2
	+ \frac{\omega g}{4} N(N-1)
  \la{Hsquare}
\eea
where $x_{j}$ are coordinates of $N$ particles, $p_{j}$ are
their canonic momenta, and $g$ is the coupling constant. We took the mass of
the particles to be unity.

The model (classical and quantum) occupies an exceptional place in
physics and mathematics and has been studied extensively \cite{Perelomov-book,PerQuant,Sutherland-book}.
hCM similarly to other Calogero-Moser systems can be obtained by the Hamiltonian reduction of the system of non-interacting Hermitian matrices moving in external harmonic potential \cite{Perelomov-book}. In this reduction the $N$ coordinates of particles $x_{j}$ appear as eigenvalues of simply evolving $N\times N$ matrix. The model  is completely integrable and its solutions can be presented in terms of the eigenvalue problem for a finite dimensional matrix (see Sec.~\ref{app-perelomov} for details).

A remarkable fact is that the hydrodynamic limit $N\to\infty$ of system (\ref{hCM}) can be found exactly using the methods of collective field theory \cite{JevickiSakita,Sakita-book,Jevicki-1992} or using the methods of \cite{2005-AbanovWiegmann,2009-AbanovBettelheimWiegmann}. The hydrodynamic Hamiltonian can be written in terms of density and velocity fields as
\bea
	H &=&\int dx\, \rho\,\left[\frac{v^2}{2}
	+ \frac{g^2}{2}\left(\pi\rho^H-\partial_x\ln\sqrt{\rho}\right)^2+\frac{\omega^2x^2}{2}\right]
 \la{Htd} \\
 	&=& \int dx\, \rho\,\frac{1}{2}\Big|v-i\omega x
	+ig(\pi\rho^H-\partial_x\ln\sqrt{\rho})\Big|^{2} +const,
\eea
where $\rho^H$ is Hilbert transform of $\rho$ defined as a principal value integral
\be
	\rho^H=\frac{1}{\pi} \dashint^{+\infty}_{-\infty}dy\; \frac{\rho(y)}{y-x}.
\ee
The density and velocity fields have a Poisson's bracket
\be
	\{\rho(x),v(y)\}=\delta'(x-y).
 \la{PBrhov}
\ee
In this work we stress that the hydrodynamic form (\ref{Htd},\ref{PBrhov}) can be used even for the finite number of particles $\int \rho(x)\,dx=N$ (see Sec.~\ref{hydro-limit}).

A goal of this paper is to find ``soliton solutions'' of the system (\ref{Htd},\ref{PBrhov}). Corresponding soliton solutions of the Calogero model without an external potential ($\omega=0$) are well known. A single solitons solution was found in \cite{1995-Polychronakos,AndricBardekJonke-1995}, and generalized to multi-soliton solutions in \cite{2009-AbanovBettelheimWiegmann}.

Let us first explain what we mean by a soliton solution. Soliton is usually defined as ``a pulse that maintains its shape while it travels at constant speed''. Obviously this definition does not make any sense in the presence of an external harmonic potential. Instead, we should talk about finite-dimensional reductions of an infinitely dimensional system (\ref{Htd},\ref{PBrhov}). Namely, if there is a solution of that system of the form $\rho(x,t)=\rho(x; \{z_{j}\})$ (and $v(x,t)=v(x; \{z_{j}\})$) so that the time dependence of  $\rho$ and $v$ is reduced to $M$ complex parameters $z_{j}(t)$ ($j=1,2,\ldots,M$) with known time dependence, we call it an $M$-soliton solution. For example, in translationally invariant systems a one-soliton solution has a form $\rho(x,t)=\rho(x-z(t))$ with $z(t)=vt$ which is consistent with the standard soliton definition.

The main result of the paper is the $M$-soliton solutions of (\ref{Htd},\ref{PBrhov}). It is presented  in Sec.~\ref{msolitonhydro}. The complex parameters $z_{j}(t)$ of this multi-soliton solution in turn satisfy a ``dual'' Calogero model (\ref{pmoty}). Therefore, the complicated dynamics of an infinite-dimensional hCM (\ref{Htd}) is reduced to an $M$-dimensional dynamics of  complex Calogero system. We have to stress here that finding an explicit solution is still a non-trivial problem as one also has to relate initial conditions $z_{j}(t=0)$ of a dual Calogero system (\ref{pmoty}) with initial density and velocity profiles of (\ref{Htd}). This is done implicitly in (\ref{TDredM},\ref{TDzdot}). The derivations used in obtaining (\ref{TDredM},\ref{TDzdot}) are very close to the ones used in \cite{2009-AbanovBettelheimWiegmann}.

Remarkably, the finite dimensional reduction can also be performed in the finite-dimensional hCM (\ref{hCM}) with $N$ particles. The evolution of (\ref{hCM}) with finely tuned initial conditions can be described as a motion of few complex parameters $z_{j}(t)$ $(j=1,2,\ldots, M)$ with $M<N$. This result is not published anywhere to the best of our knowledge and is another important result of this paper.

The organization of the paper is the following. To introduce some notations and for the reader's convenience we start with a brief review of a solution of hCM (\ref{hCM}) in Sec.~\ref{app-perelomov}. We formulate a self-dual dynamical system which can be reduced to hCM in Sec.~\ref{dualCM}. A similar self-dual system has appeared before in Ref.\cite{2009-AbanovBettelheimWiegmann} for a trigonometric Calogero model. We extend it to hCM. We show that this self-dual system allows for the reductions which correspond to soliton solutions of hCM. Several examples of such reductions are given in Sec.~\ref{sec:Msolutions}. In Sec.~\ref{sec:utau} we encode positions of hCM particles and their momenta by poles of meromorphic functions and derive equations for those functions using the approach of  \cite{2005-AbanovWiegmann,2009-AbanovBettelheimWiegmann}. We use these equations to rewrite the dynamics of hCM in hydrodynamic form in Sec.~\ref{hydro-limit} and present soliton solutions in the hydrodynamic limit in Sec.~\ref{sec:hydrosolitons}. In concluding Sec.~\ref{sec:conclusion} we discuss possible generalizations of this work and some open questions. Some details of calculations are delegated to appendices.

\section[Solution of hCM with $N$ particles]{Solution of hCM with $N$ particles}
\label{app-perelomov}

Here we briefly review the explicit solution of hCM (see Ref.\cite{Perelomov-book} for review). In this solution the coordinates of Calogero particles $x_{j}(t)$ can be found at any time as eigenvalues of a simple matrix $Q(t)$. For the sake of brevity we do not  discuss here neither a geometric meaning of the solution nor how this solution could be obtained (see \cite{Perelomov-book}). Instead, we just introduce notations and give explicit formulas that we use in this work.

Let us introduce the following $N\times N$ matrices:
\bea
	X_{ij} &=& \delta_{ij}x_i,
 \la{X-matrix} \\
	L^{\pm} &=& L \pm i\omega X, \;\;\mbox{where } L_{ij} = p_i\delta_{ij} +(1-\delta_{ij})\frac{ig}{x_i - x_j},
 \la{L-matrix} \\
	M_{ij} &=& g\left[\delta_{ij}\sum_{l=1 (l\neq i)}^{N}\frac{1}{(x_{i}-x_{l})^{2}}
	-(1-\delta_{ij})\frac{1}{(x_{i}-x_{j})^{2}}\right].
 \la{M-matrix}
 \eea
These matrices depend on time through $x_{j}(t)$ and $p_{j}(t)$ and satisfy important identities:
\bea
 	[X,L] = ig(e\otimes e^{T}-1),
 \la{momentum} \\
 	Me=0 \qquad \mbox{and}\qquad e^{T}M=0.
 \la{Me}
\eea
Here $e^{T}=(1,1,\ldots,1)$.

It is straightforward to show that the equations of motion of hCM (\ref{hCM})
\bea
    \dot{x}_{j} &=& p_{j},
 \la{csmeq1} \\
    \dot{p}_{j} &=& -\omega^{2}x-g^{2} \frac{\partial}{\partial x_{j}}
    \sum_{k=1\, (k\neq j)}^{N}\frac{1}{(x_{j}-x_{k})^{2}}
 \la{csmeq2}
\eea
are equivalent to the following matrix equations
\bea
	\dot{X} +i[M,X] &=& L,
 \la{Xdot-equation} \\
	\dot{L}+i[M,L] &=& -\omega^{2}X
 \la{Ldot-equation}
\eea
or equivalently
\be
	\dot{L}^{\pm} = -i\left[M,L^{\pm}\right] \pm i\omega L^{\pm}
 \la{LM-equation}
\ee
written in terms of $L$ and $M$ matrices usually referred to as a Lax pair.
It immediately follows from (\ref{LM-equation}) (see also Eq.~\ref{LpLmev}) that the following quantities
\be
	I_{k} = \Tr (L^{-}L^{+})^{k} =  \Tr (L^{+}L^{-})^{k}
 \la{Ik}
\ee
are integrals of motion of hCM. $I_{0} \equiv \Tr 1 = N$ is the number of particles while
\be
	I_{1} =\Tr (L^{-}L^{+})=  2{\cal H}_{hCM}
\ee
is the Hamiltonian (\ref{hCM}) itself. The higher integrals of motion $I_{k}$, $k=2,3,\ldots$  are in involution, i.e. have vanishing Poisson's bracket with each other.  The existence of a high number of conserved quantities is the result of integrability of hCM.

One can also write the solution of hCM as an eigenvalue problem of a matrix which can be explicitly constructed from the initial positions and velocities of Calogero particles. Namely, the trajectories of particles are given by eigenvalues of the following matrix
\bea
	Q(t) &=& X(0)\cos(\omega t) + \omega^{-1}L(0)\sin(\omega t).
 \la{Q-explicit}
\eea
Here the matrices $X(0)$ and $L(0)$ are constructed from initial conditions $x_{j}(0)$, $p_{j}(0)$ using definitions (\ref{X-matrix},\ref{L-matrix}).

\section{Dual Calogero system and finite-dimensional reductions}
 \la{dualCM}

In this section we consider a complexified version of hCM  (\ref{csmeq1},\ref{csmeq2}). We parametrize the complex momenta $p_{j}$ by complex numbers $z_{j}$ so that the system (\ref{csmeq1},\ref{csmeq2}) is rewritten as equations symmetric in $x_{j}$ and $z_{j}$ (see (\ref{xjdot},\ref{zjdot}) below). We refer to the obtained symmetric system as to a self-dual form of hCM.
The self-dual form of hCM (\ref{xjdot},\ref{zjdot}) makes explicit the duality between particles $x_{j}$ and excitations (parametrized by $z_{j}$) of Calogero system. It is different from the action-coordinate duality explored previously in classical Calogero systems \cite{nikita}. The self-dual system for the trigonometric Calogero-Sutherland model appeared previously in \cite{2009-AbanovBettelheimWiegmann} (see \ref{sdCSM}). It is transparent in the Hirota form (\ref{bihBO}) as a symmetry between tau-functions $\tau^{-}$ and $\tau^{+}$ (see \cite{2005-AbanovWiegmann}).

After introducing the self-dual form of hCM we consider different reductions of this system: reductions of the number of points $z_{j}$  in a dual model and a real reduction ($x_{j}$ - real).  Both of these reductions combined produce soliton solutions for an original hCM.

\subsection{Self-dual Calogero system}
 \la{sec:dcm}

Here we consider $x_{j},z_{j}$ as well as $p_{j}=\dot{x}_{j}$ and $\dot{z}_{j}$ as complex numbers.
We introduce the following dynamic system:
\bea
	\dot{x}_{j} -i\omega x_{j} & = & -ig\sum_{k=1 (k\neq j)}^{N}\frac{1}{x_{j}-x_{k}}
	+ig\sum_{n=1}^{M}\frac{1}{x_{j}-z_{n}},
 \la{xjdot} \\
   	\dot{z}_{n} -i\omega z_{n} & = & ig\sum_{m=1(m\neq n)}^{M}\frac{1}{z_{n}-z_{m}}
	-ig\sum_{j=1}^{N}\frac{1}{z_{n}-x_{j}},
 \la{zjdot}
\eea
for $x_{j}(t)$ with $j=1,2,\ldots, N$ and $z_{n}(t)$ with $n=1,2,\ldots,M$. We start with the case  $M=N$.
Let us note for future use that there is a connection between the motion of center of masses of points $x_{j}$ and $z_{n}$ obvious from (\ref{xjdot},\ref{zjdot})
\be
	\sum_{j=1}^{N}(\dot{x}_{j}-i\omega x_{j}) = \sum_{n=1}^{M}(\dot{z}_{n}-i\omega z_{n}).
 \la{cmrelation}
\ee
The system (\ref{xjdot},\ref{zjdot}) is Hamiltonian.  It can be defined by its Hamiltonian given up to an additive constant $-\omega g N(N+1)/4$ by
\bea
	{\cal H}_{hCM}
	&=& -\frac{g^{2}}{2}\sum_{j=1}^{N}\left(\sum_{k=1}^{N}\frac{1}{x_{j}-z_{k}}\right)^{2}
	-\frac{g^{2}}{2}\sum_{j=1}^{N}\left(\sum_{k=1}^{N}\frac{1}{z_{j}-x_{k}}\right)^{2}
 \la{sdhCM}\\
	&+& \frac{g^{2}}{2}\sum_{j,k=1}^{N}\left(\frac{1}{x_{j}-z_{k}}\right)^{2}
	-\frac{\omega g}{2}\sum_{j,k=1}^{N}\frac{x_{j}+z_{k}}{x_{j}-z_{k}}
 \nonumber
\eea
and by a symplectic form $\Omega =\sum_{j,k=1}^{N}S_{jk}dz_{k}\wedge dx_{j}$, where $S_{jk}=ig(x_{j}-z_{k})^{-2}$ and corresponding
Poisson's bracket $\{z_{k},x_{j}\}=(S^{-1})_{kj}$. We notice that the system (\ref{xjdot},\ref{zjdot},\ref{sdhCM}) is symmetric under simultaneous exchange $x_{j}\leftrightarrow z_{j}$ and $g\to -g$.

Equations (\ref{xjdot},\ref{zjdot}) are first order differential equations. The dynamics is fully defined by initial values of complex $x_{j}$, $z_{j}$, i.e., by $2N$ complex numbers.

Taking a time derivative of (\ref{xjdot}) (and similarly of (\ref{zjdot})) and using (\ref{xjdot},\ref{zjdot}) we exclude first time derivatives. \footnote{After excluding first derivatives one has to reorganize products of fractions to exclude $z_{j}$ from the first equation. For this purpose the following identity comes in handy $
\frac{1}{x-y}\frac{1}{x-z} + \frac{1}{y-z}\frac{1}{y-x}+\frac{1}{z-x}\frac{1}{z-y} = 0$.
} As a result we obtain the decoupled systems of second order differential equations
\bea
   \ddot{x}_{j} & = & -\frac{g^{2}}{2}\frac{\partial}{\partial x_{i}}\sum_{i\neq j}^{N}
   \frac{1}{\left(x_{i}-x_{j}\right)^{2}}-\omega^{2}x_{j}, \qquad j=1,\dots, N
 \la{pmotx}
 \\
    \ddot{z}_{j} & = & -\frac{g^{2}}{2}\frac{\partial}{\partial z_{i}}\sum_{i\neq j}^{M}\frac{1}{\left(z_{i}-z_{j}\right)^{2}}-\omega^{2}z_{j}, \qquad j=1,\dots, M.
 \la{pmoty}
\eea
The system (\ref{pmotx}) is a complex version of  the system of equations of motion obtained from hCM (\ref{hCM}), i.e., equivalent to (\ref{csmeq1},\ref{csmeq2}). We refer to a system (\ref{pmoty}) as to the Calogero system dual to (\ref{pmotx}) or simply: \textit{the dual Calogero system}. We outline the Lax formalism for this dual system and its correspondence to the one for the original system of Sec.~\ref{app-perelomov}  in \ref{sec:lf} and \ref{sec:integrals}.

As soon as initial values of $x_{j}$ and $\dot{x}_{j}$ are chosen, their evolution is defined by (\ref{pmotx}). Then the motion of complex points $z_{j}$ is, on one hand, defined by the motion of $x_{j}$ through (\ref{xjdot},\ref{zjdot}) while on the other hand they evolve as Calogero system (\ref{pmoty}). This shows that one can think of the transformation $x_{j}(t)\to z_{j}(t)$ given by (\ref{xjdot}) as of the B\"acklund transformation from one solution of (\ref{pmotx}) to the other. We do not explore the connection of our results with B\"acklund transformations further in this work.

\subsection{Reduction of number of particles in a dual system}
 \la{sec:solred}

A remarkable fact is that the derivation of (\ref{pmotx},\ref{pmoty}) from (\ref{xjdot},\ref{zjdot}) also holds if $M\neq N$ (we are interested here in $M<N$) and one can still think of (\ref{pmoty}) as of a dual system for (\ref{pmotx}) consisting of smaller number $M<N$ of particles. The difference with $M=N$ is that in the case $M<N$ one can not generically solve (\ref{xjdot}) to find $z_{n}$ for an arbitrary choice of $x_{j},\dot{x}_{j}$. Some fine tuning of initial values of $x_{j},\dot{x}_{j}$ is necessary. Instead, one can specify $N$ complex points $x_{j}$ and $M$ complex points $z_{n}$ and then find $\dot{x}_{j}$ from (\ref{xjdot}).
Then by (\ref{zjdot}) the motion of $N$ points $x_{j}$ is reduced to a motion of $M<N$ complex points $z_{n}$ governed by a dual Calogero system (\ref{pmoty}) having fewer degrees of freedom than the original system (\ref{pmotx}). We refer to this reduction as to a dimensional reduction or to an  $M$-soliton reduction of (\ref{pmotx}).

The soliton reduction can also be understood as a limit in which some coordinates of dual particles go to infinity. Indeed, let us consider the self-dual system (\ref{xjdot},\ref{zjdot})  with $M=N$. We choose initial positions $x_{j}$ arbitrarily and initial positions $z_{n}$ so that the latter are divided into two groups. The coordinates $z_{n}$ ($n=1,2,\ldots,M$) are arbitrary while the coordinates $z_{k}$ ($k=M+1,\ldots, N$) are very far away from the origin so that for $k>M$:  $|z_{k}|\gg |x_{j}|$ for any $j$ and $|z_{k}|\gg|z_{n}|$ for any $n\leq M$. We are interested in the limit $z_{k}\to\infty$ for $k>M$.  One can see that in this limit only $M$ coordinates $z_{n}$, $n\leq M$ enter the equations for $\dot{x}_{j}$ as it is written in (\ref{xjdot}) with $M<N$. The equations for $\dot{z}_{j}$ are divided in this limit into $M$ equations for $\dot{z}_{n}$ with $n\leq M$ (see (\ref{zjdot})) and to completely decoupled system of $N-M$  points $z_{k}$ ($k=M+1,\ldots N$). The latter system is not important for us while the system (\ref{xjdot},\ref{zjdot}) with $M<N$ gives an $M$-soliton reduction as the dynamics of $z_{n}$ with $n\leq M$ is given by (\ref{pmoty}) having less degrees of freedom than (\ref{pmotx}).


\subsection{Real reduction}
 \la{sec:rr}

So far we considered $x_{j}$ as complex numbers.
It is clear, however, from (\ref{pmotx}) that once initial values of $x_{j}$ and $\dot{x}_{j}$ are  chosen to be real they stay real at later times, even though $z_j$ are moving in a complex plane. Let us specify some arbitrary real values of $x_{j}(t=0)$ and $\dot{x}_{j}(t=0)$. For $M=N$ one can generically solve an algebraic system (\ref{xjdot}) ($N$ algebraic equations with $M=N$ unknowns $z_{j}$) and find corresponding initial complex $z_{j}(t=0)$ and then using (\ref{zjdot}) initial $\dot{z}_{j}(t=0)$. This procedure defines a real reduction of the complex system (\ref{xjdot},\ref{zjdot}). We can think of (\ref{xjdot},\ref{zjdot}) as an alternative way to write the system (\ref{csmeq1},\ref{csmeq2})  or equivalently (\ref{pmotx}) understanding that initial complex values of $z_{j}$ and $\dot{z}_{j}$ are not arbitrary but constrained by reality of $x_{j}$ and $\dot{x}_{j}$.

We notice here that a true symmetry between (\ref{pmotx}) and (\ref{pmoty}) exists only for complex variables $x_{j}$ and for $M=N$. Imposing reality conditions on $x_{j}, \dot{x}_{j}$ one explicitly breaks the symmetry between (real) $x_{j}$ and (complex) $z_{n}$.

\section{Soliton solutions of hCM with $N$ particles}
 \la{sec:Msolutions}

Now we consider the case when both real and soliton reductions are applied simultaneously. In this case one can take real and imaginary parts of complex equations  (\ref{xjdot}) and write the following real equations
\bea
	\omega x_{j} &=& g\sum_{k=1 (k\neq j)}^{N}\frac{1}{x_{j}-x_{k}}
	-\frac{g}{2}\sum_{n=1}^{M}\left(\frac{1}{x_{j}-z_{n}}+\frac{1}{x_{j}-\bar{z}_{n}}\right),
 \la{xjreal} \\
	p_{j} &=&
	i\frac{g}{2}\sum_{n=1}^{M}\left(\frac{1}{x_{j}-z_{n}}-\frac{1}{x_{j}-\bar{z}_{n}}\right).
 \la{pjreal}
\eea
If  $M$ complex positions $z_{n}$ are given at any time one can find both $N$ real positions $x_{j}$ and corresponding real momenta $p_{j}$. The data $x_{j},p_{j}$ are not independent but ``tuned'', i.e., related by (\ref{xjreal},\ref{pjreal}) through the values of $M$ complex parameters  $z_{n}$ ($2M$ real parameters).

Equations (\ref{xjreal}) have an electrostatic interpretation. Indeed, (\ref{xjreal}) can be obtained as extrema conditions for  the following function
\bea
	E = \sum_{j=1}^{N}\frac{\omega x_{j}^{2}}{2g} -\sum_{j<k}\ln|x_{j}-x_{k}| +\frac{1}{2}\sum_{j=1}^{N}\sum_{n=1}^{M}\ln|x_{j}-z_{n}|^{2}.
 \la{estatic}
\eea
This function coincides with  an ``electrostatic energy'' of $N$ particles with unit charges interacting through a logarithmic potential (2d Coulomb potential). The particles are restricted to move along a straight line (a real axis) and are in the presence of $2M$ external charges $-1/2$ placed at $z_{n},\bar{z}_{n}$ and an external harmonic potential.
We notice here that the solution of (\ref{xjreal}) is not necessarily a minimum of (\ref{estatic}). Soliton solutions correspond to any extremum (maximum, minimum or saddle point) of (\ref{estatic}). It is important to stress that here and in the following we choose the signs $\omega>0$ and $g>0$ which guarantees that the harmonic potential in (\ref{estatic}) is confining.

\subsection{Background}
 \la{background}

As an ultimate case of $M$-soliton reduction we consider $M=0$ which gives a static solution. Indeed, (\ref{xjreal},\ref{pjreal}) in the limit $z_{n}\to\infty$ for all $n$ becomes $p_{j}=0$ for all $j$ and coordinate of particles in equilibrium are defined by (\ref{xjreal}) as:
\be
\la{Hermite}
	\omega x_j = g \sum_{k=1 (k\neq j)}^{N} \frac{1}{x_j-x_k}.
\ee
It is well known that a solution of this system of algebraic equations is given by the roots of  $N$-th  Hermite polynomial (Stiltjes formula \cite{Szego-1975,mehta}). Namely,
\be
	x_{j}(t) = \sqrt{\frac{g}{\omega}} h_{j}, \qquad H_{N}(h_{j})=0, \;\; j=1,2,\ldots,N.
 \la{HzerosSol}
\ee

\subsection{One soliton solution}
 \la{1soliton}

Consider the case $M=1$. Equations (\ref{xjreal},\ref{pjreal}) give
\bea
   	 \omega x_j
	&=& g\sum_{k=1 (k\neq j)}^{N} \frac{1}{x_j-x_k}
	-\frac{g}{2} \left( \frac{1}{x_j-z_{1}} + \frac{1}{x_j-\bar{z}_{1}}\right),
 \la{M1xj} \\
   	p_j &=&  i\frac{g}{2} \left( \frac{1}{x_j-z_{1}} - \frac{1}{x_j-\bar{z}_{1}}\right).
 \la{M1pj}
\eea
The equations (\ref{M1xj}) can be viewed as a new generalization to the Stieltjes problem (\ref{Hermite}) (see Refs. \cite{Szego-1975,forrester,orive}). To the best of our knowledge this generalization to the Stieltjes problem has not been studied and exact solutions of (\ref{M1xj}) are not known. One can think of (\ref{M1xj}) as of definition of some polynomials $P_{N}(x,z_{1})\equiv \prod_{j}(x-x_{j})$ such that $P_{N}(x_{j},z_{1})=0$ for $j=1,2,\ldots N$. In the limit $z_{1}\to\infty$ we have $P_{N}(x,z_{1})\to H_N(x\sqrt{\omega/g})$. We make some progress in describing these solutions in the limit $N\gg1$ in Sec.\ref{sec:hydrosolitons}.

The equation (\ref{pmoty}) in the case $M=1$ takes an especially simple form
\be
	\ddot{z}_{1} = -\omega^{2}z_{1}
 \la{oscillator}
\ee
and can be easily solved
\be
	z_{1}(t) = Ae^{i\omega t} + B e^{-i\omega t},
\ee
i.e., the trajectory of $z_{1}$ is an ellipse in the complex plane. Using (\ref{cmrelation}) for $M=1$ we obtain the parameters of this ellipse
\be
	z_{1}(t) = z_{1}(0) e^{i\omega t} + \frac{\sin \omega t}{\omega}\left[P(0)-i\omega X(0)\right],
 \la{1solellipse}
\ee
where $z_{1}(0)$ is the initial position of $z_{1}$ in the complex plane and $X=\sum_{j=1}^{N}x_{j}$, $P=\sum_{j=1}^{N}p_{j}$ are the center of mass and the total momentum of the system at $t=0$. Both $X(0)$ and $P(0)$ are in turn defined by $z_{1}(0)$ through (\ref{M1xj},\ref{M1pj}).

Let us consider for simplicity a particular initial value $z_{1}(0)=i b$ with $b>0$. Then the solution of (\ref{M1xj}) gives $X=0$.\footnote{We do not know how to prove this statement. Equations (\ref{M1xj}) have a symmetry $x_{j}\to-x_{j}$ and numerical solutions suggest that this symmetry is unbroken resulting in $X=0$.} The equation of the ellipse in this case is
\be
	z(t)=ib \cos(\omega t)-(b-P(0)/\omega)\sin(\omega t),
 \la{explicitellipse}
\ee
where we find from (\ref{M1pj})
\be
	P(0)=-b\sum_{j}\frac{g}{x_{j}^{2}+b^{2}}<0.
 \la{P0micro}
\ee
The inequality means that $a=b-P(0)/\omega>b$ so that the major semiaxis $a$ of the ellipse is always along the real axis. In the limit $b\to\infty$,  $P(0)\sim -\frac{gN}{b}$ and major and minor semiaxes are $a\approx b+ \frac{gN}{\omega b}$ and $b$ respectively. The eccentricity of the ellipse goes to zero (ellipse becomes a circle) as $b\to \infty$. In the opposite limit $b\to 0$ we have $P(0)\sim -\frac{g}{b}$ giving $a\sim \frac{g}{\omega b}$. In this limit the ellipse has a large eccentricity with the major semiaxis $a\sim b^{-1}\to\infty$ as the minor semiaxis $b\to 0$.

\begin{figure}
\centerline{\includegraphics[scale=1.0]{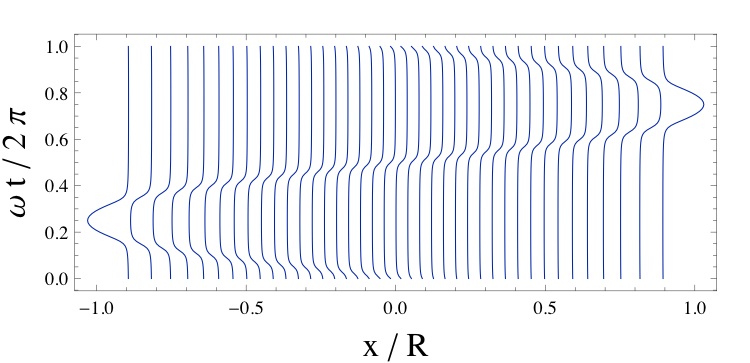}}
\caption{\label{wline}World-line diagram for a single-soliton solution of  Harmonic Calogero Model is shown for $N=40$ and the value $b\approx 0.84 < b_{c}\approx 1.67$. Lines are the world lines of individual Calogero particles. There are no crossings of world lines. However, the lump of density corresponding to a soliton travels all the way through the system. It becomes an isolated particle outside of the ``cloud'' of other particles.}
\end{figure}

Let us now fix some large value of the major semiaxis $a$ by taking $z_{1}(0)=a>0$. It is clear from this analysis that there are two different solutions of (\ref{M1xj}) corresponding to large and small values of minor semiaxis of the ellipse. These two solutions correspond to two different extrema of electrostatic energy (\ref{estatic}). For one of them all $N$ particles (``cloud'') are located around the origin, far from the external negative ``soliton'' charge placed at $a$. For the other extremum the cloud around the origin consists of $N-1$ particles. One more particle is far away from the cloud, close to the external charge. The former solution corresponds to the large minor semiaxis $b\approx a$, while the latter corresponds to $b\sim g/a$. If we decrease $a$ two corresponding values of $b$ approach each other and become equal to some ``critical'' value $b=b_{c}$. At this value the major semiaxis $a$ has a minimum value $a_{c}$. For $a<a_{c}$ there are no real solutions of (\ref{M1xj}). Later in Sec.~\ref{sec:hydrosolitons} we will show that in the limit of large $N$ this minimum occurs at $b_{c}\sim N^{1/6}$ and corresponds to a minimal major semiaxis $a_{c}-R\sim  N^{-1/6}$, where $R=\sqrt{2gN/\omega}$ is the radius of the ``cloud'' of particles. A world-line diagram of a typical single-soliton solution for $b<b_{c}$ is shown in Figure \ref{wline}. In this regime the soliton solution looks like a Newton's cradle. The soliton is essentially a single particle when its position is outside of the ``cloud''. This particle transfers its momentum all the way through the system with the other particle being kicked out from the other side of the system. Due to the interactions (in contrast to the actual Newton's cradle) the particle is dressed by other particles when inside of the cloud. This picture was qualitatively described by Polychronakos \cite{Polychronakos-2006}.

\section{Particles as poles of meromorphic functions}
 \la{sec:utau}

In this section we, following an approach of \cite{2005-AbanovWiegmann,2009-AbanovBettelheimWiegmann} consider particles of hCM as poles of meromorphic functions and derive dynamic equations satisfied by those functions.

We start by introducing two meromorphic functions $u^{\pm}(x)$ of a complex variable $x$
\bea
    u^{-}(x) = -ig\sum_{j=1}^{N}\frac{1}{x-x_{j}}+i\omega x,
    \la{u-}
  \\
    u^{+}(x) = ig\sum_{n=1}^{M}\frac{1}{x-z_{n}}.
    \la{u+}
\eea
These functions are completely defined by their poles $x_{j}$ and $z_{n}$ which move as Calogero particles (\ref{pmotx},\ref{pmoty}). The function $u^{-}(x)$ is defined by its poles -- the coordinates of hCM particles $x_{j}$. The function $u^{+}(x)$ is defined by the coordinates of the dual model $z_{n}$ or alternatively by its values at $x_{j}$ given by
\be
    u^+(x_j) = p_j- i\omega x_j + ig\sum_{k=1 (k\neq j)}^{N} \frac{1}{x_j-x_k}.
    \la{u+j}
\ee
Conditions (\ref{u+j}) are equivalent to (\ref{xjdot}). Notice that the r.h.s. of (\ref{u+j}) appears in the factorized form of the hCM Hamiltonian (\ref{Hsquare}).

Having defined $u^{\pm}(x)$ by (\ref{u-},\ref{u+}) we can rewrite the system (\ref{xjdot},\ref{zjdot}) as a single equation
\be
    u_{t}+\left[\frac{u^{2}}{2}+i\frac{g}{2}\left(u^{+}-u^{-}\right)_{x}
    +\frac{\omega^{2}x^{2}}{2}\right]_{x}=0
    \la{hBO}
\ee
with $u\equiv u^+ + u^-$. Indeed, assuming the form (\ref{u-},\ref{u+}) and taking the residues of (\ref{hBO}) at points $x_{j}$, $z_{n}$ we reproduce (\ref{xjdot},\ref{zjdot}) respectively. The equation (\ref{hBO}) is a version of a bidirectional Benjamin-Ono equation \cite{2009-AbanovBettelheimWiegmann} modified for hCM. A key advantage of (\ref{hBO}) is that the number of particles $N$ does not enter the equation explicitly and, therefore, this form is well-suited for taking hydrodynamic limit $N\to \infty$. Before discussing this limit in Sec.~\ref{hydro-limit} we also give a bilinear Hirota form of (\ref{hBO})
\be
    \left(iD_t + \frac{g}{2}D_x^2 -\omega xD_{x} -\frac{\omega}{2}\right)\tau^-\cdot \tau^+ = 0,
    \la{bihBO}
\ee
where $\tau^{\pm}$ are given by:
\bea
	u^+ = ig\, \p_x\ln\tau^+, \\
	u^- = -ig\, \p_x\ln\tau^-
\eea
and, e.g., $D_{t}f\cdot g$ denotes Hirota derivative.
We note here that up to trivial time-dependent factors tau-functions are given by
\bea
	\tau^{-}(x) &=& \prod_{j=1}^{N}(x-x_{j})
	= \det(x-Q),
 \la{tau-det}\\
	\tau^{+}(x) &=& \prod_{n=1}^{M}(x-z_{n})
	= \det(x-\hat{Q}),
 \la{tau+det}	
\eea
where the $N\times N$ matrix $Q$ is given by (\ref{Q-explicit}) and $\hat{Q}$ is the corresponding dual $M\times M$ matrix (\ref{Qhat-explicit}). The self-duality of hCM is expressed then as an obvious symmetry of (\ref{bihBO}) under the exchange of tau-functions $\tau^{-}\leftrightarrow \tau^{+}$, $g\to -g$.





\section{Equations of motion in hydrodynamic form and hydrodynamic limit}
 \la{hydro-limit}

Here we rewrite equations of motion for hCM in a hydrodynamic form for finite $N$ and then consider the hydrodynamic limit of those equations, i.e., the limit of infinitely many particles $N\to \infty$. We start with equations for $u^{\pm}(x)$ and with corresponding analyticity and reality conditions and then present the equations of motion in hydrodynamic form, i.e., written for particle density and velocity fields. We again follow the approach of \cite{2009-AbanovBettelheimWiegmann}.

Let us start by rewriting hCM (\ref{hCM}) in terms of fields $u^{\pm}$. One can show that (\ref{hCM}) is identical to
\bea
	{\cal H}_{hCM}
	&=& \frac{1}{4\pi g}\oint dz \left(\frac{u^{3}}{3} +\frac{ig}{2}u^{+}\partial_{z}u^{-}
	+\omega^{2}z^{2}u\right)
 \la{hCMu}\\
 	&=& \frac{1}{4\pi g}\oint dz \left({u^{+}}^{2}u^{-}+u^{+}{u^{-}}^{2}
	+\frac{ig}{2}u^{+}\partial_{z}u^{-} +\omega^{2}z^{2}u^{-}\right)
 \nonumber
\eea	
by using the definition (\ref{u-}) and the property (\ref{u+j}). The contour of integration in (\ref{hCMu}) goes around the real poles $x_{j}$ of $u^{-}(z)$ counter-clockwise and does not encircle any of complex poles $z_{n}$ of $u^{+}(z)$. The equations of motion (\ref{hBO}) is equivalent to (\ref{csmeq1},\ref{csmeq2}).

The poles of $u^{-}(z)$ are real and one can parameterize the real analytic function $iu^{-}(z)$ by a real function of a real variable $\rho(x)$ - a particle density field. We introduce
\be
	\rho(x) = \sum_{j=1}^{N}\delta(x-x_{j})
 \la{rho-micro}
\ee
and rewrite (\ref{u-}) as a Cauchy transform of $\rho(x)$:
\be
	u^{-}(z) =i\omega z- ig\int_{-\infty}^{+\infty} dx\, \frac{\rho(x)}{z-x},
\ee
where $z$ is a complex number not coinciding with any of poles of $u^{-}(z)$ (e.g., $\I(z)\neq 0$).

The field $u^{-}(z)$ is discontinuous on a real axis with the discontinuity related to the density of particles. More precisely
\be
 \la{u-td}
	u^-(x\pm i0)=\mp \pi g \rho + i(\pi g \rho^H +\omega x)
\ee
with the discontinuity
\be
	u^{-}(x+i0)-u^{-}(x-i0)=-2\pi g \rho(x).
 \la{u-jump}
\ee
Using (\ref{u-},\ref{u+}) as well as (\ref{xjdot},\ref{zjdot}) and (\ref{u-td},\ref{u-jump}) after some calculations we obtain that on the real axis
\be
	\rho(x)u^{+}(x) = -ig\left(\pi\rho(x)\rho^{H}(x)-\frac{1}{2}\partial_{x}\rho\right)
	+i\omega x \rho(x) +\sum_{j=1}^{N}\dot{x}_{j}\delta(x-x_{j}).
 \la{rhov1}
\ee
We identify the last term of the r.h.s as a momentum density of the system
\be
	\rho(x)v(x) =\sum_{j=1}^{N}\dot{x}_{j}\delta(x-x_{j}),
 \la{v-micro}
\ee
where $v(x)$ is the velocity field. We divide (\ref{rhov1}) by $\rho(x)$ and obtain
\be
 \la{u+td}
	u^+(x)=v-ig\left(\pi\rho^H-\partial_x\log\sqrt{\rho}\right)+i\omega x.
\ee
Equations (\ref{u-td},\ref{u+td}) give the relation between fields $u^{\pm}(x)$ and microscopic density and velocity fields. We notice here that these relations are exact even in the case of finite number of particles $N$. The density and velocity fields have a conventional Poisson's bracket (\ref{PBrhov}). Substituting (\ref{u-td},\ref{u+td}) into the Hamiltonian (\ref{hCMu}) we arrive to the Hamiltonian of hCM in a hydrodynamic form (\ref{Htd}).

Hamilton equations following from (\ref{Htd},\ref{PBrhov}) are the Euler and the continuity equations for density and velocity fields:
\bea
	\rho_t &+& \partial_x(\rho v) = 0,
 \la{Cont} \\
	v_t &+& \partial_x\left(\frac{v^2}{2}+w(\rho)+\frac{\omega^2x^2}{2}\right) = 0,
 \la{Euler}
\eea
where a chemical potential $w(\rho)$ is given by:
\be
	w(\rho)=\frac{1}{2}(\pi g \rho)^2 + \pi g^2\rho^H_x
	-\frac{g^2}{2}\frac{1}{\sqrt{\rho}}\partial_x^2\sqrt{\rho} .
\ee
We remark here that although the equations in this section are written in hydrodynamic form, they are still valid for a finite number of particles $N$ and are equivalent to the corresponding equations for hCM. In the case of finite $N$ the density and velocity fields are singular functions given by their microscopic definitions (\ref{rho-micro},\ref{v-micro}). All expressions involving these fields and their products should be properly regularized as it is explained above. The key point of the regularization is to use the definitions (\ref{u-},\ref{u+}) of $u^{\pm}$ as meromorphic functions.

Let us now go to a hydrodynamic limit. This simply means that from now on we treat $\rho(x)$ and $v(x)$ as continuous (even smooth) fields forgetting the discrete nature of hCM particles. Note, that the information about the total number of particles is still preserved in the relation $\int\rho\,dx=N$ and one should do some rescaling of fields when going to the large $N$ limit (see Sec.~\ref{1solitonhydro}).
Having specified an initial configuration $\rho(x,t=0)$, $v(x,t=0)$ one can in principle solve (\ref{Cont},\ref{Euler}) and find density and velocity fields at all times. An interesting class of solutions (multi-soliton solutions) of  (\ref{Cont},\ref{Euler}) is realized for a fine-tuned initial configurations of fields. As the number of particles $N$ and the number of poles $M$ of the field $u^{+}(x)$ are independent parameters, one can take a hydrodynamic limit $N\to\infty$ keeping $M$ finite and fixed. As a result one obtains solutions in which the dynamics of continuous fields $\rho(x,t)$ and $v(x,t)$ is reduced to a motion of $M$ points $z_{n}(t)$ in a complex plane. This is a finite-dimensional reduction of an infinitely dimensional hydrodynamic system. We refer to this reduction as to an $M$-soliton solution.

In the next section we consider several examples of soliton solutions in the large $N$ limit.

\section{Soliton solutions of hCM in hydrodynamic limit}
 \la{sec:hydrosolitons}

\subsection{Background}
 \la{bacgroundhydro}

Let us find the configuration $\rho(x)$ and $v(x)$ with given $\int\rho(x)\,dx=N$ that minimizes the energy (\ref{Htd}). We rewrite (\ref{Htd}) in a manifestly positive form
\bea
	H &=& \int dx\, \rho\,\left[\frac{v^2}{2}
	+ \frac{1}{2}\left(\pi g\rho^H-g\partial_x \log\sqrt{\rho}-\omega x\right)^2\right]+const
  \la{HtdSQ} \\
	&=& \int dx\, \rho\, \frac{1}{2}\left|u^{+}(x)\right|^{2} +const,
 \nonumber
\eea
where we used (\ref{u+td}) to obtain the last line.
It is easy see that the minimal energy condition is
\be
	u^{+}(x)=0,
\ee
or writing it separately for real and imaginary parts and using (\ref{u+td}):
\bea
	g(\pi\rho^H-\partial_x\log\sqrt{\rho})-\omega x = 0,
 \la{rhovac} \\
	v(x) = 0.
 \la{vvac}
\eea
Eq.~(\ref{rhovac}) is the hydrodynamic form of the equation (\ref{Hermite}). It describes the distribution of zeros of Hermite polynomials $H_{N}(x\sqrt{\omega/g})$. In the limit $N\rightarrow\infty$ we think of $\rho(x)$ and $v(x)$ as of continuous fields. In this limit the solution of (\ref{rhovac}) is given by a Wigner's semi-circle law
\be
	\rho_{0}(x)=\frac{\omega}{\pi g}\sqrt{R^{2}-x^{2}},
	\qquad R=\sqrt{\frac{2 g N}{\omega}}.
 \la{Wsc}
\ee
Eq.~(\ref{rhovac}) also appears in the context of Random Matrix Theory (see, for example, Refs. \cite{LECHTENFELD,itoi}). We notice here that both the density at the origin
$\rho_{0}(0)=\bar{\rho}=\frac{1}{\pi} \sqrt{\frac{2 \omega N}{g}}$ and the radius of the cloud of particles $R$ are proportional to $N^{1/2}$. The main correction to (\ref{Wsc}) in the next to leading order in $1/N$ comes from the fact that the largest zero of $H_{N}(x\sqrt{\omega/g})$ is not $R$ but is given asymptotically by $x_{max}\approx R-\gamma_{1}R^{-1/6}$, where the constant $\gamma_{1}\approx 1.8557\ldots$ is related to zeros of Airy functions. It is also notable that the distance between neighbor roots goes as $x_{n+1} - x_n\sim \bar{\rho}^{-1}\sim N^{-1/2}$ close to the origin and $x_{N}-x_{N-1} \sim R^{-\frac{1}{6}}$ near the boundary of the cloud. \cite{Szego-1975}


\subsection{One-soliton solution}
 \la{1solitonhydro}

The one-soliton solution is given by
\be
	u^{+}(x) = \frac{ig}{x-z_{1}(t)},
 \la{u+1sol}
\ee
with $z_{1}(t)$ satisfying (\ref{pmoty}) for $M=1$ or (\ref{oscillator}).
Using (\ref{u+td}) we rewrite (\ref{u+1sol}) as
\be
 \la{TDred1}
	v-i\omega x=\frac{ ig}{x-z_{1}} + ig(\pi\rho^H-\partial_x\log\sqrt{\rho}).
\ee
This relation allows one, in principle, to find density and velocity fields from the position $z_{1}$ at any moment of time. The soliton parameter $z_{1}(t)$ is moving in a complex plane along the ellipse (\ref{1solellipse}). Therefore, (\ref{TDred1}) gives a $1$-dimensional reduction of an infinite dimensional Calogero system in hydrodynamic limit defined by (\ref{Htd},\ref{PBrhov}). Eq.~(\ref{TDred1}) is a hydrodynamic analogue of (\ref{xjdot}) with $M=1$.

Taking real and imaginary parts of (\ref{TDred1}) we obtain hydrodynamic counterparts of (\ref{M1xj},\ref{M1pj})
\bea
	g(\pi\rho^H-\partial_x\log\sqrt{\rho})-\omega x
	=\frac{g}{2} \left( \frac{1}{x-z_{1}} + \frac{1}{x-\bar{z}_{1}}\right),
 \la{1solTD} \\
	v=\frac{ig}{2} \left( \frac{1}{x-z_{1}} - \frac{1}{x-\bar{z}_{1}}\right).
 \la{1solTDv}
\eea
It is remarkable that the velocity field of a one-soliton solution is given explicitly by a simple expression (\ref{1solTDv}). The equation (\ref{1solTD}) defines, albeit implicitly, the density field for a one-soliton solution. Comparing (\ref{1solTD},\ref{1solTDv}) with the corresponding background equations (\ref{rhovac},\ref{vvac}) we see that the fields for a one-soliton solution are obtained by perturbing the background configurations by terms $\sim 1/z_{1}$. In particular, in the limit $z_{1}\to\infty$ we go back to the equilibrium configuration (\ref{rhovac},\ref{vvac}). In the large $N$ limit the term $\partial_{x}\log\sqrt{\rho}$ and the right hand side of (\ref{1solTD}) are both suppresed by $1/N$ with respect to other two terms.
We also notice here that the solution of (\ref{oscillator}) for $z_{1}(t)$ is given by (\ref{1solellipse}), where
\bea
	P = \int dx\, \rho v, \qquad\qquad X = \int dx\, \rho x.
 \la{PXpar}
\eea
are the total momentum and the center of mass of the system. Of course, finding $P(0)$ and $X(0)$ from $z_{1}(0)$ using (\ref{1solTD},\ref{1solTDv}) is still a non-trivial problem.

In the limit $\omega\to 0$, $N\to\infty$ and $\bar{\rho}=\rho_{0}(0)=\frac{1}{\pi} \sqrt{\frac{2 \omega N}{g}}=const$ the equation (\ref{1solTD}) gives rise to Lorenzian
shaped solitons in agreement with solitons obtained by Polychronakos \cite{1995-Polychronakos} and Andric et.~al. \cite{AndricBardekJonke-1995} for a model without harmonic potential and with the background density $\bar{\rho}$.

As the exact solution of (\ref{M1xj},\ref{1solTD}) is not available we briefly discuss the solution in the limit of large $N$ in the next to leading order in $1/N$.

Rescaling variables $(\rho,x,z_{1},v)\rightarrow (\rho, x,z_{1},v)\sqrt{N}$  in (\ref{1solTD},\ref{1solTDv}) one can easily see that the right hand sides of (\ref{1solTD},\ref{1solTDv}) are of the order of $1/N$. Therefore, in the leading order in $N$ one has density and velocity given by (\ref{vvac},\ref{Wsc}).
%
%
%

The correction to (\ref{rhovac}) consists of two parts: the correction to the background solution without solition and to the correction caused by the presence of the soliton, i.e. by the right hand side of (\ref{1solTD}). Here we are interested only in the latter.

First, let us assume that the solution of (\ref{1solTD}) is given by a smooth function $\rho(x)$. Then we have:
\begin{eqnarray}
	\rho(x,t) & = & \rho_{0}(x)+\frac{1}{\pi}\frac{y_{1}}{(x-x_{1})^{2}+y_{1}^{2}}
 \label{eq:rsol}\\
	v(x,t) & = & -g\frac{y_{1}}{(x-x_{1})^{2}+y_{1}^{2}},
 \label{eq:vsol}
\end{eqnarray}
where we denoted $z_{1}(t)=x_{1}(t)+iy_{1}(t)$. The solution (\ref{eq:rsol},\ref{eq:vsol}) describes a lump of density of the changing width $\sim y_{1}(t)$ located at the moving point $x_{1}(t)$. The point $z_{1}(t)$ moves according to (\ref{1solellipse}). Let us start with $z_{1}(0)=ib$. Using (\ref{eq:rsol},\ref{eq:vsol}) and (\ref{PXpar}) we find the parameters $X(0)=0$ and $P(0)\approx -\frac{g}{2b}+\omega b -\omega \sqrt{R^{2}+b^{2}}$. The major semiaxis of the ellipse is given (see Sec.\ref{1soliton}) by
\be
	a= b-\omega^{-1}P(0) \approx \frac{g}{2\omega b} +\sqrt{R^{2}+b^{2}}.
 \la{semiaxisa}
\ee
As a function of time $z_{1}(t)=ib\cos\omega t -a\sin\omega t$.
We can identify several interesting limits corresponding to different values of $b$.

\paragraph{Large:  $b\gg R\sim\sqrt{N}$.} In this case we have $a\approx b +R^{2}/(2b)$. The trajectory of $z_{1}$ is close to a circle of a very large radius $b$. The width $b$ of the soliton is bigger than the size of the cloud. In this case there is no pronounced lump of the density. Instead the whole cloud oscillates slightly around the origin.

\paragraph{Intermediate I: $N^{1/6}\ll b\ll \sqrt{N}$.} For $b\ll R$ we obtain from (\ref{semiaxisa}) $a\approx R+\frac{g}{2\omega b} +\frac{b^{2}}{2R}$. The major semiaxis $a$ has a minimum at $b_{c}\approx (gR/2\omega)^{1/3}\sim  N^{1/6} \sqrt{g/2\omega}$.  The width of the soliton is $b$ at $t=0$. It is much smaller than the size of the system and one can see a very well pronounced lump of density while soliton travels through the system. The width of the soliton somewhat changes in time but remains much larger than an interparticle distance inside the cloud. Therefore, the continuous approximation is still valid in this regime at all times. The soliton in this regime is a well-pronounced lump of density which oscillates inside the cloud of particles.

\begin{figure}
\centerline{\includegraphics[scale=0.5]{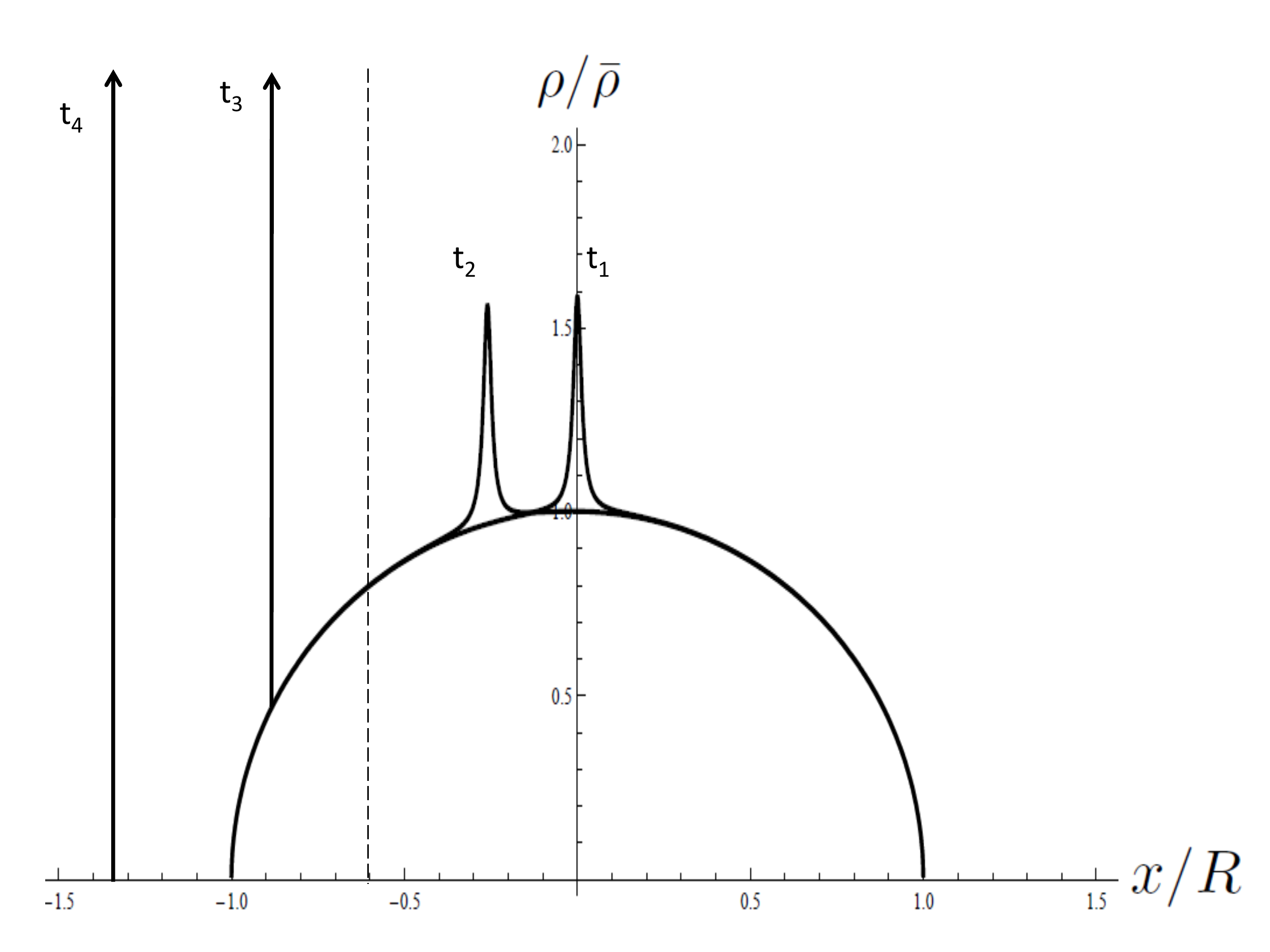}}
	\caption{\label{hydro} Time evolution ($t_{1}<t_{2}<t_{3}<t_{4}$) of a one-soliton solution in
	hCM is shown schematically in the large $N$ limit. The figure corresponds to the regime Intermediate II. As soliton moves to the left its width decreases and becomes of the order of interparticle distance at some $x$ (shown by the dashed line). After this point the continuum approximation is not valid and soliton is represented by the delta-function (shown by an arrow). This figure is a continuous analogue of Figure \ref{wline}}
\end{figure}

\paragraph{Intermediate II: $N^{-1/2}\ll b\ll N^{1/6}$.} This is, probably, the most interesting regime. For an initial configuration the width of the soliton $b$ is still much bigger than an interparticle distance $N^{-1/2}$ in the middle of the harmonic trap. Therefore, we still can use a continuous approximation and the value $a\approx R+\frac{g}{2\omega b} +\frac{b^{2}}{2R}$. However, as a function of time $y_{1}(t)$ decreases and at some point becomes of the order of an interparticle distance at the point $x_{1}(t)$. \footnote{A maximal interparticle distance is at the edge of the cloud and is of the order $N^{-1/6}$. \cite{Szego-1975}} Starting from this time we cannot use the continuous approximation. Instead, we assume that the density can be divided into a delta function corresponding to a single particle plus a continuous background with $N-1$ particles. In the limit when $y_{1}$ is much smaller than an interparticle distance we have simply
\be
	\rho(x,t) = \rho_{0}(x) +\delta(x-x_{1}(t)).
 \la{rho+delt}
\ee
The evolution of density in this case is shown in Figure~\ref{hydro} and is a continuous analogue of a world-line diagram for finite number of particles shown in Figure~\ref{wline}. Notice, that in this regime a boundary particle is kicked out of the cloud and travels outside of the cloud for some fraction of the period of the motion.

\paragraph{Small: $b\ll N^{-1/2}$.} In this limit the continuous approximation is invalid already at $t=0$ and we consider an isolated particle at the origin with other $N-1$ particles forming a continuous cloud (\ref{rho+delt}) for all times. The value of $P(0)$ is given by a microscopic formula (\ref{P0micro}) which is dominated in this case by a particle at the origin $P(0)\approx -g/b$. It gives $a\approx b + g/(\omega b)$. The density evolution is given by a particle moving in the semicircle background (\ref{rho+delt},\ref{Wsc}).

\subsection{Multi-soliton solution}
 \la{msolitonhydro}

Here we briefly list equations describing $M$-soliton solutions of (\ref{Cont},\ref{Euler}). The density and velocity fields are completely defined by complex coordinates $z_{n}$ through (\ref{u+}) which can be rewritten using (\ref{u+td}) as
\be
 \la{TDredM}
	v-i\omega x= ig\sum_{k}\frac{1}{x-z_k} + ig(\pi\rho^H-\partial_x\log\sqrt{\rho}).
\ee
An initial configuration of $M$ complex numbers $z_{n}$ defines initial velocity and density fields through (\ref{TDredM}). After density and velocity fields are found one can easily determine initial $z$-velocities using a hydrodynamic analogue of (\ref{zjdot}) which has a form
\be
 \la{TDzdot}
	\dot{z}_n - i\omega z_n
	= ig\sum_{m=1 (m\neq n)}^{M} \frac{1}{z_n-z_m} + \pi g (\rho +i \rho^H).
\ee
After initial velocities $\dot{z}_{n}$ are obtained, the dynamics of $z_{n}$ is defined by (\ref{pmoty}) so that $z_{n}$ can be found as eigenvalues of the matrix $\hat{Q}$ (\ref{Qhat-explicit}).

The dynamical problem of finding a multi-soliton solution of (\ref{Cont},\ref{Euler}) is reduced to finding  the density $\rho(x,\{z_{n}\})$ from (\ref{TDredM}). The latter is still  complicated, albeit time-independent, integral equation.

\section{Conclusion}
 \la{sec:conclusion}

In this work we used a self-dual formulation (\ref{xjdot},\ref{zjdot}) of a harmonic Calogero system (hCM) (\ref{hCM}) to find an $M$-soliton reduction of an hCM with $N$ particles $M<N$. Soliton solutions can be obtained by fine tuning the initial conditions $x_{j}$, $\dot{x}_{j}$ for Calogero particles. We found a hydrodynamic formulation of this reduction and then took a hydrodynamic limit $N\to\infty$ keeping $M$ finite. As a result we obtained an $M$-soliton solution of an infinitely-dimensional hydrodynamic system.

%

The derivations of this article are based on similar derivations of \cite{2009-AbanovBettelheimWiegmann} made for Calogero-Sutherland model. We emphasized in this work that the soliton reductions are possible even for finite number of particles while Ref.~\cite{2009-AbanovBettelheimWiegmann} considers only finite soliton solutions of an infinite-dimensional translationally invariant model. We gave the generalizations of finite dimensional reduction to the cases of trigonometric and rational Calogero-Moser systems in \ref{sdCSM} and \ref{sdCM} respectively. The generalization of finite-dimensional reductions to the elliptic Calogero models is also rather straightforward and will be given elsewhere. We also expect that the generalizations of our results to more general Calogero-Moser systems related to Lie algebras are possible.

In this work we did not discuss the meaning of the presented soliton reductions of hCM neither within the projection method of solving hCM (see \cite{Perelomov-book,1978-KKS}) nor within inverse scattering formalism \cite{1980-Krichever}. It is interesting to find the corresponding descriptions of both self-dual formulation of hCM and of the soliton reductions.

The self-duality of hCM given by (\ref{xjdot},\ref{zjdot}) and used in this work is different from the known dualities of Calogero models \cite{Ruijsenaars-1994, nikita}. It would be interesting to have a precise relation between those dualities as well as the connection with the known bispectral property of Calogero-Moser systems \cite{wilson,kasman,dui}.

hCM and many other Calogero models remain integrable after quantization. Moreover, many results  obtained for classical Calogero models have direct analogues for their quantum counterparts. In particular, the pole ansatz can be extended to the quantum case \cite{2005-AbanovWiegmann}. The classical soliton solutions of Calogero models correspond to quasi-particle excitations of the corresponding quantum models \cite{Sutherland-book}. It would be interesting to give quantum analogues of the results presented here.

\section{Acknowledgments}

We are grateful to E. Bettelheim,  I. Krichever, A. Lamacraft, N. Nekrasov, L. Takhtajan, and J. Verbaarschot for useful discussions.
 The work of A. G. A. was supported by the NSF under Grant No. DMR-0906866.

\appendix


\section[\hspace{2.5cm}Lax formalism for dual CM]{Lax formalism for a dual Calogero system}
 \la{sec:lf}

In this Appendix we describe Lax matrices for a dual Calogero system. The formalism is almost identical to the one presented in Sec.~\ref{app-perelomov} for an original hCM. In addition we introduce an intertwining matrix $F$ relating corresponding matrices between dual systems.

Let us define matrices dual to (\ref{X-matrix},\ref{L-matrix},\ref{M-matrix}) as
\bea
	\hat{Z}_{mn} &=& \delta_{mn}z_m(t),
 \la{Z-matrix} \\
	\hat{L}^{\pm} &=& \hat{L} \pm i\omega \hat{Z}, \;\;\mbox{where }
	\hat{L}_{mn} = \dot{z}_{m}\delta_{mn}+(1-\delta_{mn})\frac{ig}{z_m - z_n},
 \la{Lhat-matrix} \\
	\hat{M}_{mn} &=& g\left[\delta_{mn}\sum_{l=1 (l\neq m)}^{M}\frac{1}{(z_{m}-z_{l})^{2}}
	-(1-\delta_{mn})\frac{1}{(z_{m}-z_{n})^{2}}\right].
 \la{Mhat-matrix}
\eea
Similarly to (\ref{momentum},\ref{Me}) we have:
\bea
 	[\hat{Z},\hat{L}] = ig(e_{M}\otimes e_{M}^{T}-1_{M}),
 \la{hatmomentum} \\
 	\hat{M}e_{M}=0 \qquad \mbox{and}\qquad e_{M}^{T}\hat{M}=0.
 \la{hatMe}
\eea
Here $e_{M}^{T}=(1,1,\ldots,1)$ is a row vector made out of $M$ ones and $1_{M}$ is a unit $M\times M$ matrix. To avoid confusion we will denote the vector $e$ from (\ref{Me}) by $e_{N}$ here.

It is obvious that other formulas of Sec.~\ref{app-perelomov} can also be written in terms of dual variables and matrices. For example, similarly to $x_{j}(t)$ the values of $z_{n}(t)$ at any time can be found as eigenvalues of a $\hat{Q}$ matrix
\bea
	\hat{Q}(t) &=& \hat{Z}(0)\cos(\omega t) + \omega^{-1}\hat{L}(0)\sin(\omega t).
 \la{Qhat-explicit}
\eea

The dual variables $z_{j}$ are related to original variables $x_{j}$ through (\ref{xjdot},\ref{zjdot}) and the natural question is how the corresponding matrices and, in particular, integrals of motion for the dual system are related to the ones for an original system. Here we relate the matrices (\ref{X-matrix},\ref{L-matrix}) to (\ref{Z-matrix},\ref{Lhat-matrix}). Then in \ref{sec:integrals} we find the relations between corresponding integrals of motion.

In the following we consider matrices  (\ref{X-matrix},\ref{L-matrix},\ref{Z-matrix},\ref{Lhat-matrix}) as functions of parameters $x_{j},z_{n}$ only, with time derivatives expressed in terms of these parameters using (\ref{xjdot},\ref{zjdot}).
We introduce one more ``intertwining'' rectangular matrix $F$ of the size $N\times M$
\be
	F_{in} = \frac{ig}{x_{i}-z_{n}}.
 \la{F-matrix}
\ee
The matrix $F$ depends on both direct and dual variables and provides a connection between dual systems as we will see below.
It is straightforward to show that the following identity holds for both upper and lower signs
\be
	L^{\pm}F =F \hat{L}^{\pm} -\omega g(1\pm 1) e_{N}\otimes e_{M}^{T},
 \la{LLhat}
\ee
where multiplication is the matrix multiplication. We also find the identities
\bea
    \la{eN}
	L^{-}e_{N} &=& Fe_{M},
 \\
     \la{eNdag}
	e_{N}^{T}F &=&  e_{M}^{T}\hat{L}^{-},
\eea
which are equivalent to equations of motion (\ref{xjdot},\ref{zjdot}).

We conclude this Appendix by stating that the matrix $F$ obeys a simple matrix evolution equation
\be
 \la{Fdot}
	\dot{F} = -i\omega F - iMF + iF\hat{M},
\ee
while $\hat{L}^{\pm}$ satisfy equations fully analogous to (\ref{LM-equation})
\be
	\dot{\hat{L}}^{\pm} = -i\left[\hat{M},\hat{L}^{\pm}\right] \pm i\omega \hat{L}^{\pm}.
 \la{LMhat-equation}
\ee
A derivation of (\ref{Fdot}) is based on (\ref{xjdot},\ref{zjdot}) and is rather straightforward albeit  somewhat cumbersome.

\section[\hspace{2.5cm}Integrals of motion]{Integrals of motion}
\la{sec:integrals}

It follows from (\ref{LM-equation}) that
\be
	\partial_{t}(L^{-}L^{+}) = -i\left[M,L^{-}L^{+}\right]
 \la{LpLmev}
\ee
and, therefore, (\ref{Ik}) are integrals of motion. In fact, the time evolution (\ref{LpLmev}) does not change eigenvalues of $L^{-}L^{+}$ and describes an isospectral deformation of this matrix. Similar conclusion can be derived for matrices $\hat{L}^{-}\hat{L}^{+}$ and $\hat{L}^{+}\hat{L}^{-}$ using (\ref{LMhat-equation}).
Here we relate the integrals of motion (\ref{Ik}) to analogous expressions for the dual system.

Let us start with an easily verifiable identity
$$
	[\hat{L}^{+},\hat{L}^{-}]=2i\omega [\hat{Z},\hat{L}]=-2\omega g (e_{M}\otimes e_{M}^{T}-1_{M}).
$$
We proceed as
\bea
	F\hat{L}^{+}\hat{L}^{-}
	&=& F\hat{L}^{-}\hat{L}^{+}-2\omega g F(e_{M}\otimes e_{M}^{T}-1_{M})
 \nonumber \\
 	&=& L^{-}\Big(L^{+}F+2\omega g e_{N}\otimes e_{M}^{T}\Big)
	-2\omega g F(e_{M}\otimes e_{M}^{T}-1_{M})
 \nonumber \\
 	&=& L^{-}L^{+}F+2\omega g F,
 \la{Fcomm}
\eea
where we used (\ref{LLhat}) and (\ref{eN}).
If $f$ is an eigenvector of $\hat{L}^+\hat{L^{-}}$ and $\lambda$ is a corresponding eigenvalue i.e., $\hat{L}^{+}\hat{L}^{-}f=\lambda f$, it follows from (\ref{Fcomm}) that $L^{-}L^{+}$ has an eigenvalue $\lambda +2\omega g$ with the corresponding eigenvector $Ff$. We conclude that $M$ eigenvalues of $\hat{L}^{-}\hat{L}^{+}$ are identical (after the shift by $2\omega g$) to the $M$ eigenvalues of $L^{+}L^{-}$. We show below that the remaining $N-M$ eigenvalues of $L^{+}L^{-}$ are constants given by (\ref{tevs}). Therefore, integrals of motion of the original and dual hCM are simply related.

We start with the relation between the integrals of motion of dual systems for the case $M=N$. In this case the matrix $F$ is square and invertible (we assume that $z_{j}\neq x_{k}$ for any $j,k=1,\ldots N$). Then one can find the matrices $\hat{L}^{\pm}$ for the dual system from (\ref{LLhat}) etc.
In particular,  (\ref{Fcomm}) can be written as
\be
	L^-L^+=F(\hat{L}^+\hat{L^{-}} - 2\omega g 1_N)F^{-1}.
\ee
One immediately concludes that integrals of motion of dual systems are connected by a very simple relation
\be
	I_{k} = Tr(L^-L^+)^k = Tr(\hat{L}^+\hat{L^{-}} - 2\omega g 1_{N})^k.
\ee

To consider the case $M<N$ we exploit the fact that the dimensional reduction to $M$-soliton solution can be obtained by taking some of $z_{j}$ to infinity as it is described in Sec.~\ref{sec:solred}. We divide $z_{j}$ into two groups. We keep $z_{1},\ldots,z_{M}$ finite and take $z_{M+k}\equiv \tilde{z}_{k}$ for $k=1,\ldots (N-M)$ to infinity. We take this limit for the matrix $\hat{L}^+\hat{L^{-}}=(\hat{L}^{-})^{2}+2i\omega \hat{Z}\hat{L}^{-}$ and leave only non-vanishing matrix elements. We use the fact that all $x_{j}$ are chosen to be finite. The matrix obtained in the limit has a block-triangular form and its eigenvalues are given by the eigenvalues of $(\hat{L}^+\hat{L^{-}})_{M}$ reduced to the size $M\times M$ and to the eigenvalues of the $(N-M)\times (N-M)$ matrix $2\omega g B$ defined as
\be
	B_{ij} = \left(N-M-\sum_{k=1 (k\neq i}^{N-M}\frac{\tilde{z}_{i}}{\tilde{z}_{i}-\tilde{z}_{k}} \right)\delta_{ij}
	+(1-\delta_{ij})\frac{\tilde{z}_{i}}{\tilde{z}_{i}-\tilde{z}_{j}}
\ee
It is easy to show\footnote{The matrix $B$ is triangular in the basis of $f^{(k)}$, $k=0,1,\ldots,(N-M-1)$ defined by $(f^{(k)})_{i}=(\tilde{z}_{i})^{k}$.} that eigenvalues of $B$ are $1,2,3,\ldots, N-M$.
Therefore, the first $N-M$ eigenvalues of $L^-L^+$ are trivial and given by
\be
 \la{tevs}
 	\lambda_{s}=2\omega g s, \qquad s=0,1,2,\ldots, (N-M-1).
\ee
 The remaining $M$ eigenvalues are not trivial and coincide with those of the $M\times M$ matrix of the dual model $\hat{L}^{+}\hat{L}^{-}$ shifted by $2\omega g$. This fact illustrates the meaning of $M$-dimensional reduction for integrals of motion. In particular, for the background solution $M=0$ all $N$ eigenvalues of $L^{-}L^{+}$ are given by $2\omega g s$, $s=0,1,2,\ldots (N-1)$. The latter result is known and can be obtained directly from the properties of Hermite polynomials (see eqs.~10a,b of Ref.~\cite{LPLM}).

\section[\hspace{2.5cm}Solitons for Sutherland model]{Solitons as finite dimensional reductions of N-particle Sutherland Model}
\la{sdCSM}

Here, for the sake of completeness we give a self-dual form of the Calogero-Sutherland model (trigonometric case of Calogero model) as it appeared in \cite{2009-AbanovBettelheimWiegmann}. Then we give formulas for soliton reductions.

Calogero-Sutherland Model describes particles on a circle interacting with inverse sine-squared (chord-distance) interactions
\begin{equation}
	H=\frac{1}{2}\sum_{j=1}^{N}p_{j}^{2}+\frac{1}{2}\left(\frac{\pi}{L}\right)^{2}
	\sum_{j,k=1;j\neq k}^{N}\frac{g^{2}}{\sin^{2}\left[\frac{\pi}{L}(x_{i}-x_{j})\right]},
\end{equation}
where $L$ is the circumference of the circle.
Positions and momenta of particles on a circle can be characterized by $w_{j}=e^{\frac{2\pi i x_{j}}{L}}$ and $p_{j}=-i (L/2\pi) \dot{w}_{j}/w_{j}$,  where $0\leq x_{j}<L$.

The self-dual form of the Sutherland Model analogous to (\ref{xjdot},\ref{zjdot}) is:
\bea
	i\frac{\dot{w}_{j}}{w_{j}} &=& \frac{g}{2}\left(\frac{2\pi}{L}\right)^{2}
	\left(\sum_{k=1}^{M}\frac{w_{j}+u_{k}}{w_{j}-u_{k}}-\sum_{k=1}^{N}
	\frac{w_{j}+w_{k}}{w_{j}-w_{k}}\right),
 \la{wjdot} \\
	-i\frac{\dot{u}_{j}}{u_{j}} &=& \frac{g}{2}\left(\frac{2\pi}{L}\right)^{2}
	\left(\sum_{k=1}^{N}\frac{u_{j}+w_{k}}{u_{j}-w_{k}}
	-\sum_{k=1}^{M}\frac{u_{j}+u_{k}}{u_{j}-u_{k}}\right)
 \la{ujdot}
\eea
for $M=N$.
Here the ``positions of solitons'' are labeled by complex numbers $u_{j}$ with $|u_{j}|\neq 1$.
The finite dimensional reduction of the Sutherland model, i.e. $M$-soliton solutions are given by (\ref{wjdot},\ref{ujdot}) with $M<N$.

Taking real and imaginary parts of (\ref{wjdot}) we obtain the following relations between soliton positions and positions and momenta of particles:
\bea
	&& \sum_{k=1}^{N}\frac{w_{j}+w_{k}}{w_{j}-w_{k}}+\frac{1}{2}
	\sum_{k=1}^{M}\left(\frac{w_{j}+u_{k}}{w_{j}-u_{k}}+\frac{w_{j}
	+\frac{1}{\bar{u}_{k}}}{w_{j}-\frac{1}{\bar{u}_{k}}}\right)=0,
 \\
	&& p_{j}=-\frac{\pi g}{2L}\sum_{k=1}^{M}\left(\frac{w_{j}+u_{k}}{w_{j}-u_{k}}-\frac{w_{j}
	+\frac{1}{\bar{u}_{k}}}{w_{j}-\frac{1}{\bar{u}_{k}}}\right).
\eea
Here we used that $\bar{w}_{j}=w_{j}^{-1}$ for particles on a circle.
The static solution is obtained for $M=0$. It is easy to check that up to translation it is given by $p_{j}=0$, $x_{j}=j L/N$ (or $w_{j}=e^{i\frac{2\pi}{N}j}$).

\section[\hspace{2.5cm}Solitons for Calogero model]{Soliton reduction of Calogero model (rational case)}
\la{sdCM}

Here we discuss how the soliton reduction can be implemented for the rational Calogero-Moser system or Calogero model (CM). This model is given by Hamiltonian (\ref{hCM}) with $\omega=0$. It can be written in a self-dual form by taking $\omega = 0$ in (\ref{xjdot},\ref{zjdot}). Then an $M$-soliton reduction can be obtained by taking $M<N$ in (\ref{xjdot},\ref{zjdot}). Although this reduction is well defined for a complexified system, applying it to the original real Calogero model we run into the following difficulty.  The real equations (\ref{xjreal}) do not have solutions for $M<N$ if $\omega=0$. It is easy to understand from the electrostatic interpretation. Indeed, it is not possible to keep $N$ repelling charges within some finite interval on a line with the using the negative charge $M<N$ in the absence of an additional harmonic potential (see (\ref{estatic}) with $\omega=0$).
We show here how to overcome this difficulty and obtain an $M$-soliton reduction for CM.

Let us  consider the following change of variables
\bea
	x_j' &=&  \frac{x_j}{\cos(\omega t)},
 \la{CMtrans1} \\
	\omega t'  &=&  \tan(\omega t).
 \la{CMtrans2}
\eea
It is known that this transformation ``removes harmonic potential'' \cite{Perelomov-book}. Namely, if $x_{j}(t)$ is a solution of hCM, the transformed functions $x_{j}'(t')$ defined by (\ref{CMtrans1},\ref{CMtrans2}) give a solution of the Calogero model.

It is clear that an $M$-soliton reduction of hCM gives through the change of variables (\ref{CMtrans1},\ref{CMtrans2}) a corresponding reduction of CM.

Let us apply the change of variables (\ref{CMtrans1},\ref{CMtrans2}) to the self-dual form of hCM (\ref{xjdot},\ref{zjdot}). We obtain
\bea
	\dot{x}'_{j} - \frac{i\omega x'_{j}}{1+i\omega t'} & = & -ig\sum_{k=1 (k\neq j)}^{N}\frac{1}{x'_{j}-x'_{k}}
	+ig\sum_{n=1}^{M}\frac{1}{x'_{j}-z'_{n}},
 \la{CMxjdot} \\
   	\dot{z}'_{n} - \frac{i\omega z'_{n}}{1+i\omega t'} & = & ig\sum_{m=1(m\neq n)}^{M}\frac{1}{z'_{n}-z'_{m}}
	-ig\sum_{j=1}^{N}\frac{1}{z'_{n}-x'_{j}},
 \la{CMzjdot}
\eea
where we also changed $z_{n}\to z_{n}'$ similarly to (\ref{CMtrans1}).
We consider (\ref{CMxjdot},\ref{CMzjdot}) as a modified or deformed self-dual form of CM. $\omega$ is just a parameter of the deformation (there is no time scale $\omega$ in CM). At the value $\omega=0$ equations (\ref{CMxjdot},\ref{CMzjdot}) give an unmodified self-dual form of CM. At $\omega=0$ there are no real solutions $x_{j}(t)$ for $M<N$ as it was explained above. However, for $\omega\neq 0$ one obtains all soliton reductions corresponding to the ones for hCM. The obtained soliton solutions will have an explicit time dependence additional to the time-dependence of parameters $z_{n}(t)$.

Before giving an example of the reduction we stress that excluding $z_{n}$'s from (\ref{CMxjdot},\ref{CMzjdot}) one arrives to the system of second order differential equations for CM. The parameter $\omega$ does not enter these equations. Similarly, excluding $x_{j}$'s one finds that the parameters $z_{n}(t)$ form a dual CM, that is also move according to CM equations.

Let us consider the simplest example of soliton solutions for CM. Namely, we consider $M=0$-soliton reduction of rational CM corresponding to a static (background) solution of hCM (\ref{HzerosSol}). This solution is mapped to
\be
 \la{CMbckgr}
	x'_j(t')=\sqrt{\frac{g}{\omega}} h_j\sqrt{1+\omega^2t'^2}.
\ee
This equation gives $0$-dimensional reduction of CM system. It is easy to check that, indeed, (\ref{CMbckgr}) solves (\ref{CMxjdot}) for $M=0$.  The parameter $\omega$ enters the initial conditions ($t'=0$) of (\ref{CMbckgr}) and defines the time scale. The limit $\omega\to 0$ is singular and does not correspond to a physical solution of CM. In this Appendix we showed that soliton reduction of the the rational Calogero model can be implemented via mapping soliton solutions of hCM onto solutions of CM using (\ref{CMtrans1},\ref{CMtrans2}). The same kind of mapping can be done between two hCM with different frequencies which will result in new soliton solutions that will have additional explicit time dependence.

\section*{References}


\end{document}